\documentstyle[aps,prd,psfig]{revtex}
\begin{document}
\renewcommand{\thesection}{\arabic{section}}
\renewcommand{\thesubsection}{\arabic{subsection}}
\title{Chiral Properties of QCD Vacuum in Magnetars- A Nambu-Jona-Lasinio Model
with Semi-Classical Approximation} 
\author{ Sutapa Ghosh, Soma Mandal and Somenath Chakrabarty
{\thanks{E-Mail: somenath@klyuniv.ernet.in}}}
\address{Department of Physics, University of Kalyani, Kalyani 741 235,
India }
\maketitle
\noindent PACS:97.60.Jd, 97.60.-s, 75.25.+z 
\begin{abstract}
The breaking of chiral symmetry of light quarks at zero temperature in presence
of strong quantizing magnetic field is studied using Nambu-Jona-Lasinio (NJL) 
model with Thomas-Fermi type semi-classical formalism. It is found that the 
dynamically generated light quark mass can never become zero if the Landau 
levels are populated and the mass of light quarks increases with the increase 
of magnetic field strength.
\end{abstract}
\section{Introduction}
The theoretical investigation of properties of compact stellar objects in 
presence of strong quantizing magnetic field have gotten a new life after the 
recent discovery of a few magnetars \cite{R1,R2,R3,R4}. These stellar objects 
are believed to be the strongly magnetized young neutron stars. The surface magnetic
fields are observed to be $\geq 10^{15}$G. Then it is quite possible that the 
fields at the core region may go up to $10^{18}$G. The exact source of this 
strong magnetic field is of course yet to be known. These objects are
also supposed to be the possible sources of anomalous X-ray and soft gamma 
emissions (AXP and SGR). If the magnetic field is really so strong, in 
particular at the core region, they must affect most of the important physical 
properties of such stellar objects and also some of the physical
processes, e.g., the rates / cross-sections of 
elementary processes, in particular the  weak and the  electromagnetic decays / reactions taking 
place at the core region.

The strong magnetic field affects the equation of state of dense neutron
star matter. As a consequence the gross-properties of neutron stars 
\cite{R5,R6,R7,R8}, e.g., mass-radius relation, moment of inertia,
rotational frequency etc. should change significantly. In the case of
compact neutron stars, the phase transition from neutron matter to quark 
matter which may occur at the core region is also affected by strong quantizing
magnetic field.  It has been shown that a first order phase transition initiated
by the nucleation of quark matter droplets is absolutely forbidden if the 
magnetic field strength $\sim 10^{15}$G at the core region \cite{R9,R10}. 
However, a second order phase transition is allowed, provided the magnetic 
field strength $<10^{20}$G.   This is of course too high to achieve at the core
region.The study of time evolution of nascent quark matter, produced at
the core region through some higher order phase transition, shows that
in presence of strong magnetic field it is absolutely impossible to
achieve chemical equilibrium ($\beta$-equilibrium) configuration among
the constituents of the quark phase if the magnetic field strength is as
low as $ B_m
\sim 10^{14}G$.

The elementary processes, in particular, the weak and the electromagnetic
decays/reactions taking place at the core region of a neutron star are strongly 
affected by such ultra-strong magnetic fields \cite{R11,R12}. Since the cooling 
of neutron stars are mainly controlled by neutrino/anti-neutrino emission, the 
presence of strong quantizing magnetic field should affect the thermal history 
of strongly magnetized neutron stars.  Further, the electrical conductivity of 
neutron star matter which directly controls the evolution of neutron star 
magnetic field will also change significantly \cite{R12}.

Similar to the study of quark-hadron deconfinement transition inside neutron 
star core in presence of strong quantizing magnetic field, a thorough 
investigations 
have also been done on the effect of ultra-strong magnetic field on chiral 
symmetry breaking. In those studies,  quantum field theoretic formalisms
were 
mainly used \cite{R13,R14,R15,R16,R17,R18,R19}. In reference \cite{Rina} 
Inagaki et al have studied the chiral symmetry violation with NJL
model using quantum field theoretic approach in presence of strong
quantizing magnetic field. In many of these papers, the effect of
curvature with or without external magnetic field on chiral symmetry
violation have been investigated. In the studies by Gusynin et al 
in \cite{R13,VP2},
have thoroughly investigated the chiral symmetry breaking in presence of
strong external quantizing magnetic field. They have used NJL model in $2+1$ and
also in $3+1$ dimensions. It has been shown that the external magnetic field
acts as a catalyst to generate fermion mass dynamically. In the first
paper \cite{R13} they have studied it in $2+1$ dimension and showed how the
external magnetic field generated dynamical mass of fermion and broke the
dynamical flavor symmetry. They have further shown by using NJL model that 
chiral symmetry breaks dynamically even if the attractive interaction between
the fermions is extremely weak. In the second paper \cite{VP2} they
have extended the calculation to $3+1$ dimension. 
In the very nice piece of work  by \cite{R17} Lee et al have studied the
breaking of chiral symmetry for fermions in presence of external magnetic field.
It has also been shown in this work that the symmetry is broken dynamically and
further the effect of finite density and the
temperature of the system on the chiral properties of the fermions have
been investigated thoroughly  in this paper in presence of strong
magnetic field. It has been reported in this paper that there 
exists a critical density (or chemical potential) above which the chiral 
symmetry is again restored (which actually indicates the restoration of chiral 
symmetry at high enough density) and if it is treated as a chiral phase 
transition, the order will be of first order in nature. On the other hand the 
chiral symmetry is again restored at high temperature above some critical
value. In this case the transition is of second order in nature.
In an extensive review work \cite{SPK}, Klevansky has reported the dynamical 
chiral symmetry breaking in presence of strong external quantizing magnetic 
field using NJL model in $SU(2)$ and $SU(3)$ flavor space. In this paper the 
effect of density (i.e., finite chemical potential) and temperature of the 
system on chiral symmetry restoration have also been reviewed.

In the present chapter we shall study the effect of strong quantizing magnetic
field on the chiral properties of QCD vacuum with 
the help of NJL model following a semi-classical Thomas-Fermi type mean field 
approach in presence of strong quantizing external QED magnetic field. Now in NJL model, there is no in-built mechanism of color 
confinement, however, it can produce two chirally distinct phases- appropriate 
for confined quark matter within the bag and the matter outside
the bag. These phases are also known as the Wigner phase and spontaneously 
broken chiral phase respectively. Therefore, if one re-formulates the NJL model
in presence of strong quantizing magnetic field, it is quite possible to obtain
the effect of quantizing magnetic field on these two chirally distinct phases 
and hence obtain the effect of magnetic field on chiral symmetry breaking. 
Further, it is also possible to obtain bag pressure from the difference of 
vacuum energy densities of these two phases and hence its variation with strong
magnetic field. Assuming that the confinement and spontaneously broken chiral 
symmetry are synonymous, Bhaduri et. al. obtained some estimate of bag constant 
from the difference of energy densities \cite{R20} for the conventional
case.  In the present chapter we 
shall modify these original calculations of Bhaduri et. al. \cite{R20} and 
Provid${\hat{\rm{e}}}$ncia et. al. \cite{R21} to study the breaking of chiral 
symmetry of light quarks in presence of strong magnetic fields and show that 
the chiral symmetry always remains broken in presence of strong quantizing 
magnetic field if the Landau levels for quarks are populated. Our 
motivation in this work was to study the effect of strong quantizing
magnetic field on two chirally distinct phases and then obtain the
vacuum pressure as a function of strong external magnetic field.
Unfortunately, we have noticed that Wigner phase does not exist if the
Landau levels of quarks are populated and in this formalism, there is no way, 
either by controlling chemical potential i.e., the density of matter (which is 
meaningless in our investigation since we have considered QCD vacuum
state) or temperature of the system to restore chiral
symmetry. Our study is now  
basically an application of the  formalism developed  recently to study the 
equation of state of dense fermionic matter of astrophysical interest in 
presence of strong quantizing magnetic field \cite{R22}.

\section{Basic Formalism}
We start with the density matrix $\rho(x,x^\prime)$, defined by
\begin{equation}
\rho(x,x^\prime)=\sum_{{\rm{spin}},p} \psi(x)\psi^\dagger(x^\prime) 
\theta(\Lambda-\mid p_z\mid)
\end{equation}
where $ \psi $ and $ \psi^\dagger $ are respectively the negative energy Dirac 
spinor and the corresponding adjoint, satisfy the equation
\begin{equation}
h\psi=E_-\psi
\end{equation}
(and similarly for $ \psi^\dagger $) with the single particle Hamiltonian
\begin{equation}
h=\gamma_5 \vec \Sigma. (\vec p-q_f \vec A) +\beta m
\end{equation}
with
\begin{equation}
\vec \Sigma=\left ( \begin{array}{lr} \vec \sigma  & 0 \\ 0  & \vec \sigma\\
\end{array}
\right ),
\end{equation}
$\gamma_5$ and $\beta$ are the usual Dirac matrices, $\Lambda$ is the
ultra-violet cut off in the momentum integral over $p_z$ ( since we are
considering vacuum, unlike a many body fermionic statistical system we
have to put the cut off by hand) and $\vec A$ is
the electromagnetic field three vector corresponding to the external
constant magnetic field of strength $B_m$ along $z$-axis.  Here the light 
quark mass $m$ is assumed to be generated dynamically. Now in presence of 
strong quantizing magnetic field along $z$-direction, the up and down spin 
negative energy spinor solutions are therefore given by
\begin{equation}
\psi(x)=\frac{1}{(L_yL_z)^{1/2}}\exp[i(E_\nu t-p_yy-p_zz)]v_-^{(\uparrow,
\downarrow)}
\end{equation}
where
\begin{equation}
v_-^{(\uparrow)}=\frac{1}{[2E_-(E_--m)]^{1/2}}\left ( \begin{array}{c}
p_zI_\nu\\ -i(2\nu q_fB_m)^{1/2} I_{\nu-1}\\ (E_--m)I_\nu \\ 0 \end{array}
\right )
\end{equation}
and
\begin{equation}
v_-^{(\downarrow)}=\frac{1}{[2E_-(E_--m)]^{1/2}}\left ( \begin{array}{c}
i(2\nu q_fB_m)^{1/2} I_\nu\\ -p_zI_{\nu-1}\\ 0\\
(E_--m)I_{\nu-1} \end{array}
\right )
\end{equation}
where $E_-=-(p_z^2+m^2+2\nu q_fB_m)^{1/2}=-E_\nu$, is the single particle 
energy eigen value, $\nu=0,1,2...$, are the Landau quantum numbers, $q_f$ is 
the magnitude of the charge carried by $f$th flavor and
\begin{equation}
I_\nu=\left (\frac{q_fB_m}{\pi}\right )^{1/4}\frac{1}{(\nu !)^{1/2}}2^{-\nu/2}
\exp \left [{-\frac{1}{2}q_fB_m\left (x-\frac{p_y}{q_fB_m} \right )^2}\right 
] H_\nu \left [(q_fB_m)^{1/2}\left (x-\frac{p_y}{q_fB_m} \right) \right ]
\end{equation}
with $H_\nu$ is the well known Hermite polynomial of order $\nu$, and $L_y$, 
$L_z$ are respectively length scales along $Y$ and $Z$ directions. Now it can 
very easily be shown that $\nu=0$ state is singly degenerate, whereas all other 
states are doubly degenerate. We now express the density matrix, as the modified
version of Wigner transform in presence of strong quantizing magnetic field, 
in the following form:
\begin{equation}
\rho(x,x^\prime)=\sum \rho(x,x^\prime,p_y,p_z,\nu)
\exp[i\{(t-t^\prime)E_-- (y-y^\prime)p_y- (z-z^\prime)p_z\}]
\end{equation}
where the sum is over the momentum components $p_y$, $p_z$ and the Landau 
quantum number $\nu$. Since the momentum variables are continuous, the sum over
momentum components will be replaced by the corresponding integrals.
Now we have from eqn.(9)
\begin{equation}
\rho(x,x^\prime,p_y,p_z,\nu)=\sum_{spin=-1/2}^{+1/2}v(x,p_y,p_z,\nu)
v^\dagger(x^\prime,p_y,p_z,\nu)
\end{equation}

Then  substituting the negative energy up and down spinors states, we have
\begin{equation}
\rho(x,x^\prime,p_y,p_z,\nu)=\frac{1}{2E_-}[E_-
A-p_z\gamma_z\gamma_0A +m\gamma_0 A-p_\perp \gamma_y\gamma_0 B]
\theta(\Lambda-\mid p_z\mid)
\end{equation}
 (see Appendix for detail derivation)
 where the matrices $A$ and $B$ are given by
\begin{equation}
A=\left ( \begin{array}{l c c r}I_\nu I_\nu^\prime &0&0&0 \\
0 & I_{\nu-1}I_{\nu-1}^\prime &0 &0 \\
0 & 0 &I_\nu I_\nu^\prime &0 \\
0 & 0 & 0 &I_{\nu-1}I_{\nu-1}^\prime \\
\end{array}
\right )
\end{equation}
\begin{equation}
B= \left ( \begin{array}{l c c r}I_{\nu-1} I_\nu^\prime &0&0&0 \\
0 & I_\nu I_{\nu-1}^\prime &0 &0 \\
0 & 0 &I_{\nu-1} I_\nu^\prime &0 \\
0 & 0 & 0 &I_\nu I_{\nu-1}^\prime \\
\end{array}
\right )
\end{equation}
where the primes indicate the functions of $x'$. Now in the evaluation of
vacuum energy, we have noticed that it would be more convenient to define a
quantity $\mu_f$, similar to the chemical potential for the $f$th flavor
in a multi-quark statistical system in presence of strong quantizing magnetic 
field (strictly speaking we are not considering a multi-quark
statistical system and $\mu_f$ is therefore not the quark chemical
potential. However, its minimum value should be $m$ and not zero, i.e.,
in this simplified model, just like dynamical mass $m$, this quantity is
also treated as a parameter and we evaluate numerically $ \mu_f$ and $m$
and then obtain the upper limit of Landau quantum number $[\nu_{max}]$
and the cut of $\Lambda$). Then it is very easy to write
\begin{equation}
\Lambda=\left ({\mu_f}^2-m^2-2\nu q_f B_m \right)^{1/2}
\end{equation}
Since $\Lambda>0$, it is also possible to express the upper limit of $\nu$,
which is the maximum value of Landau quantum number of the levels occupied by 
$f$th flavor,  and is given by
\begin{equation}
\nu_{\rm{max}}^{(f)}=\left [ \frac{\mu_f^2-m^2}{2q_fB_m}\right ]
\end{equation}
where $[~]$ indicates the nearest integer but less than the actual number.
Now to obtain the energy density of the vacuum, we consider the NJL (chiral) 
Hamiltonian, given by
\begin{eqnarray}
H&=& \sum_{i=i}^N t(i)+\frac{1}{2} \sum_{i\neq j} V(i,j)\\
&=& \sum_{i=1}^N\gamma_5(i)\vec \Sigma(i).(\vec p_i- q_f\vec
A)-\frac{1}{2} (2g)
\sum_{i\neq j} \delta(\vec x_i-\vec x_j)[\beta(i)\beta(j)- \beta(i)
\gamma_5(i) \beta(j) \gamma_5(j)]
\end{eqnarray}
where $\Sigma$, $\gamma_5$ and $\beta$ are usual $4\times 4$ matrices
and $2g$ is the effective coupling. Here we have used the formulation of
da Providencia et al \cite{R21} of the mean field density matrix to describe 
the Dirac vacuum, thereby employing the Thomas-Fermi semi-classical method
instead of formal field theory. As we have noticed,
the Physics of condensation energy is more transparent in this method
than the formal field theoretic technique.
Assuming the magnetic field $B_m$ along $z$-direction and is constant, we can 
choose the gauge $A^\mu\equiv (0,0,xB_m,0)$. The energy of the vacuum is then 
given by
\begin{equation}
\epsilon_v=\sum_{p_{1_z},\nu_1}\int dx_1 tr_1[\{\gamma_5\vec \Sigma.
(\vec p_1-q_f \vec A)\}\rho_{p_1}]+ \epsilon_v^{(I)} 
\end{equation}
where $\rho_{p_1}$ is given by eqn.(11), the first term is the kinetic
energy part  and $\epsilon_v^{(I)}$ indicates
the interaction term, including the exchange interaction. To evaluate the 
vacuum energy, we first calculate the kinetic energy term in eqn.(18). This quantity is
proportional to the trace defined as ${\rm{Tr}}(\rho h)$, can easily be evaluated by 
using $\rho$ from eqn.(11) and the single particle Hamiltonian $h$ from 
eqn.(17). Now using the orthonormality relations for the Hermite polynomials at
the time of evaluation of integral over $dx$ and also using the anti-commutation
relations for $\gamma$-matrices, we have the first term at zero temperature
(see Appendix)
\begin{equation}
\epsilon_v^{(0)}=2N_c\sum_{f=u}^d\frac{q_fB_m}{2\pi^2}
\sum_{\nu=0}^{[\nu_{\rm{max}}^{(f)}]} (2-\delta_{\nu 0})
\int_0^\Lambda  dp_z E
\end{equation}
where $\vec p^2=p_z^2+2\nu q_fB_m$, $N_c=3$, the number of colors, and $E_-
=-E_\nu$.

In the evaluation of all the traces in this paper we have used the
following important relation:
\begin{equation}
{\rm{Tr}}(\gamma^\mu \gamma^\nu A_1A_2..B_1B_2..)=Tr(A_1A_2..B_1B_2..)
g^{\mu\nu},
\end{equation}
\begin{equation}
{\rm{Tr}}(\gamma^\mu\gamma^\nu\gamma^\lambda\gamma^\sigma
A_1A_2..B_1B_2..)=Tr(A_1A_2..B_1B_2..)(g^{\mu \nu}
g^{\sigma \lambda}-g^{\mu \lambda}g^{\nu \sigma}+g^{\mu \sigma}g^{\nu
\lambda}), 
\end{equation}
${\rm{Tr}}$(product of odd $\gamma$s with  $A$ and/or $B)=0$  etc. The other
interesting aspects of $A$ and $B$ matrices are:\\
i) $k_{1\mu}k^{2\mu}{\rm{Tr}}(A_1A_2)= (E_1E_2-k_{1z}k_{2z}){\rm{Tr}}(A_1A_2)$\\
ii) $k_{1\mu}k^{2\mu}{\rm{Tr}}(B_1B_2)= \vec k_{1\perp}.\vec
k_{2\perp}{\rm{Tr}}(B_1B_2)$\\
iii) $k_{1\mu}k^{2\mu}{\rm{Tr}}(A_1B_2)= k_{1\mu}k^{2\mu}{\rm{Tr}}(B_1A_2)=0$\\
iv) $p_{1\mu}k^{1\mu}p_{2\nu}k^{2\nu}{\rm{Tr}}(A_1B_2)\neq 0=
(E_{\nu_1}E_{\nu_2^\prime}-p_{1z}k_{1z})\vec p_{2\perp}.\vec k_{2\perp}
{\rm{Tr}}(A_1B_2)$\\
These set of relations are very recently obtained by us \cite{R22}.
Since $\gamma$ matrices are traceless and both $A$ and $B$ matrices are 
diagonal with identical blocks, it is very easy to evaluate the above traces of 
the product of $\gamma$-matrices multiplied with any number of $A$ and/or $B$, 
from any side with any order. 

To evaluate the interaction term, we first consider the direct part which is
proportional to $Tr(\beta \rho_{p_1})Tr(\beta\rho_{p_2})$ and it is very
easy to show that $Tr(\beta \gamma_5\rho)=0$ (see Appendix). 
Then using the orthonormality 
relations for Hermite polynomials and the anti-commutation relations for the 
$\gamma$-matrices, we have the direct term
\begin{equation}
V_{\rm{dir}}=-4gm^2[{\cal{V}}(\Lambda, m)]^2
\end{equation}
(The four fermion coupling is included in ${\cal{V}}$, it is $\sim
V(1,2)\rho_1\rho_2$)
where
\begin{equation}
{\cal{V}}(\Lambda,m)=\frac{N_c}{2\pi^2} \sum_{f=u}^d e_fB_m 
\sum_{\nu=0}^{[\nu_{\rm{max}}^{(f)}]} (2-\delta_{\nu 0}) \int_0^\Lambda
\frac{dp_z}{(p_z^2+m{_\nu,f}^2)^2}
\end{equation}
where $m_{\nu,f}=(m^2+2\nu q_f B_m)^{1/2}$ (see Appendix for derivation).

To evaluate the exchange term, we first calculate $Tr((\beta\rho_{p_1})
(\beta\rho_{p_2}))$. Now
\begin{equation}
\beta\rho=\frac{1}{2E_-}[E_-\beta A+p_zA\gamma_z +mA -p_\perp
B\gamma_y]
\end{equation}
   Then to obtain the trace of the product of $\beta\rho_1$ and
   $\beta\rho_2$ it is very easy to show that only the direct product  terms are
   non-zero whereas cross product terms do not contribute.
   Therefore
   \begin{eqnarray}
   \beta\rho_1 \beta\rho_2 & =& \frac{1}{4E_1E_2}[E_1\beta
   A+p_{1z}A\gamma_z+mA-p_{1\perp}B\gamma_y]\nonumber\\ &&
   [E_2\beta
   A^\prime+p_{2z}A^\prime\gamma_z+mA^\prime-p_{2\perp}B^\prime\gamma_y]
   \end{eqnarray}
   
Then using(the orthonormality relations for Hermite polynomials) at the
time of integration over $dx_1$ and $dx_2$, 
 the above trace reduces to
   \begin{equation}
   \int_{-\infty}^{+\infty}(\beta\rho_1 \beta\rho_2)dx
   =\frac{1}{4E_1E_2}[4E_1E_2+4p_{1z}p_{2z}+4m^2]
=\left [ 1+\frac{p_{1z}p_{2z}}{E_1E_2}+\frac{m^2}{E_1E_2}\right ]
\end{equation}
where both $E_1$ and $E_2$ are negative. Then in the energy contribution, after
integrating over $p_{1z}$ and $p_{2z}$, the first term gives 
\begin{equation}
\left (
\frac{N_c}{2\pi^2}\sum_{f=u}^dq_fB_m\sum_{\nu=0}^{[\nu_{\rm{{max}}}^{(f)}]}
(2-\delta_{\nu 0}) \Lambda \right
)^2
\end{equation}
Similarly the contribution from second term is given by
\begin{equation}
\left (
\frac{N_c}{2\pi^2}\sum_{f=u}^dq_fB_m\sum_{\nu=0}^{[\nu_{\rm{max}}^{(f)}]} (2-\delta_{\nu 0})
(\Lambda^2+m_{\nu,f}^2)^{1/2}\right)^2
\end{equation}
and finally, the third term is given by
\begin{equation}
m^2\left ( \frac{N_c}{2\pi^2}\sum_{f=u}^d
q_fB_m\sum_{\nu=0}^{[\nu_{\rm{max}}^{(f)}]} (2-\delta_{\nu 0})
\ln\left [ \frac{\Lambda +(\Lambda^2+m_{\nu,f}^2)^{1/2}}{m_{\nu,f}}\right ]\right)^2
\end{equation}
(see Appendix for derivation of these expressions)

To obtain the next term in the exchange part, we evaluate the trace
$Tr((\beta\gamma_5\rho_{p_1})(\beta\gamma_5\rho_{p_2}))$, which unlike
the direct case, gives non-zero contribution. Using the anti-commutation 
relations of $\gamma$-matrices and as usual with the help of orthonormality 
relations for Hermite polynomials, we finally arrive to the following
result 
\begin{equation}
-\left [ 1+\frac{p_{1z}p_{2z}}{E_1E_2}+ \frac{m^2}{E_1E_2} +m\left (
\frac{1}{E_1} +\frac{1}{E_2} \right ) \right ]
\end{equation}
(see Appendix for derivation)
The contribution to the interaction energy will again be obtained if we
integrate over $p_{1z}$ and $p_{2z}$(done in similar manner as have been
done for direct case). Then the first term is given by
\begin{equation}
\left (
\frac{N_c}{2\pi^2}\sum_{f=u}^dq_fB_m\sum_{\nu=0}^{[\nu_{\rm{max}}^{(f)}]} (2-\delta_{\nu 0})
\Lambda\right)^2
\end{equation}
The second term is given by
\begin{equation}
\left (
\frac{N_c}{2\pi^2}\sum_{f=u}^dq_fB_m\sum_{\nu=0}^{[\nu_{\rm{max}}^{(f)}]} (2-\delta_{\nu 0})
(\Lambda^2+m_{\nu,f}^2)^{1/2}\right)^2
\end{equation}
The third term is given by
\begin{equation}
m^2\left (
\frac{N_c}{2\pi^2}\sum_{f=u}^dq_fB_m\sum_{\nu=0}^{[\nu_{\rm{max}}^{(f)}]} (2-\delta_{\nu 0})
\ln\left [ \frac{\Lambda +(\Lambda^2+m_{\nu,f}^2)^{1/2}}{m_{\nu,f}}\right ]\right)^2
\end{equation}
and finally the fourth and fifth terms, which  are identical, given by
\begin{equation}
m\left (
\frac{N_c}{2\pi^2}\sum_{f=u}^dq_fB_m\sum_{\nu=0}^{[\nu_{\rm{max}}^{(f)}]} (2-\delta_{\nu 0})
\ln\left [ \frac{\Lambda +(\Lambda^2+m_{\nu,f}^2)^{1/2}}{m_{\nu,f}}\right ]\right)
\left (
\frac{N_c}{2\pi^2}\sum_{f=u}^dq_fB_m\sum_{\nu=0}^{[\nu_{\rm{max}}^{(f)}]}(2-\delta_{\nu
0}
\Lambda\right)
\end{equation}

Then combining all these terms we finally obtain the vacuum energy
density. Since the mass $m$, which is assumed to be same for both $u$
and $d$ quarks, is generated dynamically, we obtain this quantity by
minimizing the total vacuum energy density with respect to $m$, i.e.,
by putting $d\epsilon_v/dm=0$. Simplifying this non-linear equation, we
finally get
\begin{equation}
\frac{d\epsilon_v}{dm}=-P+2gQR=0
\end{equation}
where
\begin{eqnarray}
P&=&\frac{N_c}{2\pi^2}\sum_{f=u}^dq_fB_m\sum_{\nu=0}^{[\nu_{\rm{max}}^{(f)}]}(2-\delta_{\nu 0})\left [
 \frac{2 m^3 \Lambda}{m_{\nu,f}^2 }\frac{1}{(\Lambda^2+m_{\nu,}^2)^{1/2}}
-2mX\right]\\
Q&=&\frac{N_c}{2\pi^2}\sum_{f=u}^dq_fB_m\sum_{\nu=0}^{[\nu_{\rm{max}}^{(f)}]}(2-\delta_{\nu 0})\left [X-
\frac{m^2}{m_{\nu,f}^2} \frac{\Lambda}{(\Lambda^2+m_{\nu,f}^2)^{1/2}} \right ]\\
R&=&\frac{N_c}{2\pi^2}\sum_{f=u}^dq_fB_m\sum_{\nu=0}^{[\nu_{\rm{max}}^{(f)}]}(2-\delta_{\nu 0})\left [\Lambda- 4mX
\right ]
\end{eqnarray}
with
\begin{equation}
X=\ln\left [ \frac{\Lambda +(\Lambda^2+m_{\nu,f}^2)^{1/2}}{m_{\nu,f}}\right ]
\end{equation}

It is therefore obvious from eqn.(35) that the trivial solution $m=0$ is
not possible in this particular situation, or in other wards, the gap
equation  given by 
\begin{equation}
m=4g{\cal{V}} m 
\end{equation}
can not exist.  On the other hand in a non-magnetic case, or for the
magnetic field strength less than the quantum critical value, eqn.(35)
reduces to the gap equation as written above (eqn.(40)). Here
${\cal{V}}$ is the overall contribution of interaction terms. Hence
it is obvious that $m=0$, the trivial solution exists in this
non-magnetic or the conventional scenario, investigated by Bhaduri et.
al. \cite{R20}. The phase with $m=0$ is the Wigner phase and $m\neq 0$ is the 
so called Goldstone phase. Now eqn.(40) further gives
\begin{equation}
4g{\cal{V}}=1
\end{equation}
which is nothing but the well known gap equation used in BCS theory. The
gap equation therefore does not exist in presence of strong quantizing
magnetic field if the Landau levels for $u$ and $d$ quarks are
populated.

\section{Conclusion}
The non-existence of trivial solution ($m=0$) indicates the spontaneously broken
chiral symmetry in presence of strong quantizing magnetic field. Therefore 
as soon as the Landau levels are populated for light quarks in
presence of strong external magnetic field, the chiral symmetry gets broken,
the quarks become massive and the mass $m$ (assumed to be same for both
$u$ and $d$ quarks) is generated dynamically. 

Therefore we may conclude here that the Wigner phase does not exist in the 
case of relativistic Landau dia-magnetic system. Further, if the deconfinement 
transition and restoration of chiral symmetry occur simultaneously, or
in other wards, if the chiral symmetry remains restored within the bag,
as it is generally assumed, then it 
puts a big question mark whether the idea of bag model is applicable at all in
presence of strong quantizing magnetic field. Questions may also arise, that
if Wigner phase still exists inside the bag, then whether the external
quantizing magnetic field can penetrate the bag boundary, if not, what is
the underlying physics which prevents the external magnetic field from
entering the periphery of the bag.

Now to illustrate the variation of dynamical quark mass with magnetic field, we 
consider the relation
\begin{equation}
m_\pi^2=-\frac{m_0}{f_\pi^2}<\psi \bar \psi>
\end{equation}
where $m_\pi$ is the pion mass, $m_0$ is the quark current mass and $f_\pi$ is 
the pion decay constant. Using the spinor solutions given by eqns.(6) and (7) 
we get
\begin{equation}
m_\pi^2=\frac{2m_0m}{f_\pi^2}~\frac{N_c}{2\pi^2}\sum_{f=u}^d \sum_{\nu
=0}^{[\nu_{\rm{max}}^{(f)}]} (2-\delta_{\nu 0}) \ln\left
[\frac{\Lambda+(\Lambda^2+m_{\nu,f}^2)^{1/2}}{m_{\nu,f}}\right ]
\end{equation}
We have now solved eqns.(35) and (43) numerically to obtain $\Lambda$ and 
$m$ for various values of magnetic field strength. In the actual numerical work
we have solved self-consistently for the dynamical mass $m$ and the parameter 
$\mu_f$ using eqn.(15) for $\nu_{\rm{max}}^f$ and then from eqn.(14) we
get the infra red momentum cut off $\Lambda$. In our calculation we have
always used $\mu_f$ instead of $\Lambda$ which allows us to obtain
$\nu{\rm{max}}^f$. So we can not compare our result with those obtained
with zero chemical potential, since in our calculation it is just a parameter, 
it has no
resemblances with chemical potential in a finite density quark ($u,d$) matter 
system. We were forced to make this trick to obtain $m$, $\Lambda$ and
also $\nu_{\rm{max}}^f$ self-consistently from the numerical solutions
of eqns.(35) and (43). In doing numerical calculations, we have considered the 
following sets
of numerical values for the parameters. The current quark mass $m_0=10$MeV, 
pion mass $m_\pi=140$MeV, pion decay constant $f_\pi=93$MeV, coupling constant 
$g=10$GeV$^{-2}$ and electron mass $m_e=0.5$MeV. In fig.(1) we shown the 
variation of dynamically generated quark mass with the strength of magnetic 
field. As it is evident that the dynamical quark mass never goes to zero and 
diverges beyond $B_m\approx 10^{17}$G.
\begin{figure}
\psfig{figure=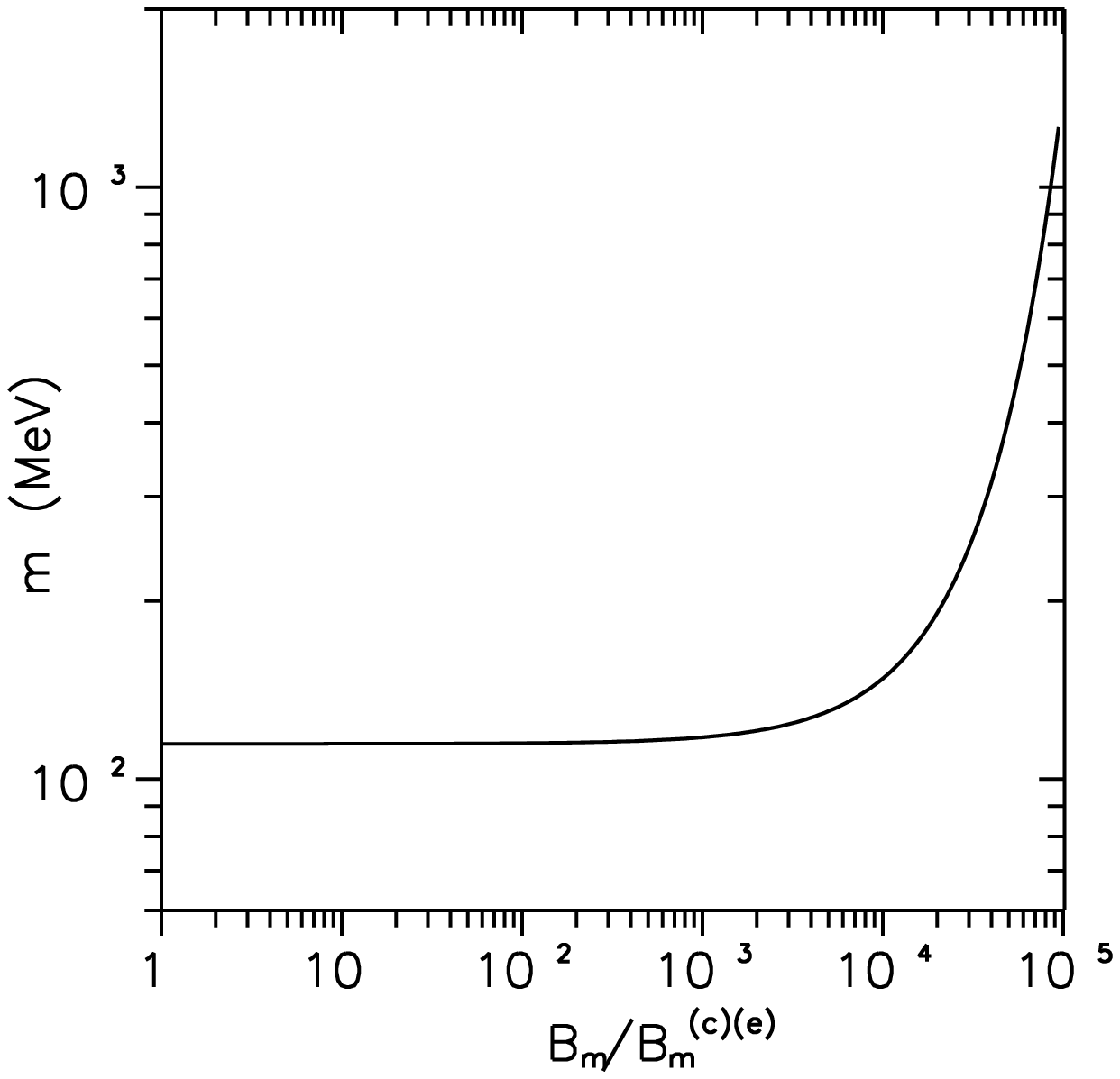,height=0.5\linewidth}
\caption{The variation of dynamically generated quark mass with the strength of
magnetic field (expressed in terms of $B_m^{(c)(e)}=4.4\times 10^{13}$G.)}
\end{figure}

\appendix
\section{}

\noindent {\bf{Evaluation Of Density Matrix:}}

To obtain the density matrix (eqn.(11)), we use
 eqns.(6) and (7), then
\begin{equation}
v^\uparrow v^{\uparrow\dagger}=\frac{1}{2E(E-m)}\left(\begin{array}{c}
p_zI_\nu\\ -i(2\nu q_fB_m)^{1/2}I_{\nu-1}\\ (E-m)I_\nu\\ 0 \end{array}
\right)
\left(
p_zI_\nu^\prime~~ i(2\nu q_fB_m)^{1/2}I_{\nu-1}^\prime~~ (E-m)I_\nu^\prime~~ 0 
\right)
\end{equation}
\begin{eqnarray}
&=&\frac{1}{2E(E-m)}\times \nonumber \\  &&\left(\begin{array}{lccr}
p_z^2I_\nu I_\nu^\prime & ip_z(2\nu q_fB_m)^{1/2}I_\nu I_{\nu-1}^\prime & 
p_z(E-m)I_\nu I_\nu^\prime & 0 \\
-ip_z(2\nu q_fB_m)^{1/2}I_{\nu-1}I_\nu^\prime & 2\nu
q_fB_mI_{\nu-1}I_{\nu-1}^\prime & -i(2\nu
q_fB_m)^{1/2}(E-m)I_{\nu-1}I_\nu^\prime & 0 \\
p_z(E-m)I_\nu I_\nu^\prime & i(2\nu q_fB_m)^{1/2} (E-m)I_\nu
I_{\nu-1}^\prime & (E-m)^2 I_\nu I_\nu^\prime & 0 \\
0 & 0 &0&0
\end{array}
\right)
\end{eqnarray}
and
\begin{equation}
v^\uparrow v^{\uparrow\dagger}=\frac{1}{2E(E-m)}\left(\begin{array}{c}
i(2\nu q_fB_m)^{1/2}I_{\nu-1}\\ -p_z I_{\nu-1}\\ 0\\
(E-m)I_{\nu-1}\end{array}\right)
\left(-i(2\nu q_fB_m)^{1/2}I_\nu~~-p_z
I_{\nu-1}^\prime~~0~~(E-m)I_{\nu-1}^\prime \right)
\end{equation}
\begin{eqnarray}
&=&\frac{1}{2E(E-m)}\times \nonumber \\ &&
\left(\begin{array}{lccr}
(2\nu q_fB_m)I_\nu I_\nu^\prime & - ip_z(2\nu q_fB_m)^{1/2}
I_\nu I_{\nu-1}^\prime & 
  0 & i(2\nu q_fB_m)^{1/2}(E-m)I_\nu I_{\nu-1}^\prime\\
+ip_z(2\nu q_fB_m)^{1/2}I_{\nu-1}I_\nu^\prime & p_z^2
I_{\nu-1}I_{\nu-1}^\prime & 0 & -(E-m)p_z I_{\nu-1}I_{\nu-1}^\prime\\ 
0 & 0 & 0 & 0\\ 
-i(2\nu q_fB_m)^{1/2}(E-m)I_{\nu-1}I_\nu^\prime &
-p_z(E-m)I_{\nu-1}I_{\nu-1}^\prime & 0 & (E-m)^2 I_{\nu-1}
I_{\nu-1}^\prime \\
\end{array}\right)
\end{eqnarray}
Adding 
\begin{eqnarray}
&&v^\uparrow v^{\uparrow\dagger}+v^\downarrow v^{\downarrow\dagger}
=\frac{1}{2E}\times \nonumber \\ &&\left(\begin{array}{lccr}
(E+m)I_\nu I_\nu^\prime & 0 & p_zI_\nu I_\nu^\prime & 
   i(2\nu q_fB_m)^{1/2}I_\nu I_{\nu-1}^\prime\\
0 & (E+m)
I_{\nu-1}I_{\nu-1}^\prime & -i(2\nu q_fB_m)^{1/2}  I_{\nu-1}I_\nu^\prime &
-p_z I_{\nu-1} I_{\nu-1}^\prime\\ 
p_z I_\nu I_\nu^\prime & i(2\nu q_fB_m)^{1/2} I_\nu I_{\nu-1}^\prime & (E-m)
I_\nu I_\nu^\prime & 0\\ 
-i(2\nu q_fB_m)^{1/2}(E-m)I_{\nu-1}I_\nu^\prime &
-p_zI_{\nu-1}I_{\nu-1}^\prime & 0 & (E-m) I_{\nu-1}
I_{\nu-1}^\prime \\
\end{array}\right)
\end{eqnarray}
which may easily be simplified after a little algebra and reduces to 
\begin{equation}
\rho(x,x^\prime,p_y,p_z,\nu)=\frac{1}{2E_-}[E_-
A-p_z\gamma_z\gamma_0A +m\gamma_0 A-p_\perp \gamma_y\gamma_0 B]
\theta(\Lambda-\mid p_z\mid)
\end{equation}
where the matrices $A$ and $B$ are given by eqns.(12) and (13).

\noindent {\bf{Evaluation of Kinetic Term:}}

Free Hamiltonian
\begin{equation}
h=\gamma^5 \vec\Sigma.\vec p +\beta m
\end{equation}
substituting $\Sigma$ from eqn.(4) and using the properties of $\gamma $
-matrices we get
\begin{equation}
h=\vec\alpha.\vec p +\beta m
\end{equation}
the usual one. With the gauge choice $A^\mu=(0,0,xB_m,0)$, we have
\begin{equation}
h=\gamma_5 \vec\Sigma.(\vec p-q_f\vec A)+\beta m
\end{equation}
by the same technique  as above, we have 
\begin{equation}
h=\vec\alpha.(\vec p-q_f\vec A)+\beta m
\end{equation}
Now to evaluate ${\rm{Tr}}(h\rho)$, where
\begin{equation}
h=(\vec\gamma.(\vec p-q_f\vec A)+m)\gamma_0
\end{equation}
we take $A_y=q_fxB_m $, whereas all other components are zero. Then we can
express the above Hamiltonian in the following form
\begin{equation}
h=(\gamma_x p_x+\gamma_y(p_y-q_fxB_m)+\gamma_zp_z+m)\gamma_0
\end{equation}
substituting
\begin{equation}
(q_fB_m)^{1/2}\left(\frac{p_y}{q_fB_m}-x\right)=-\zeta
\end{equation}
we have
\begin{equation}
h=(\gamma_x p_x-(q_fB_m)^{1/2}\gamma_y\zeta+\gamma_zp_z+m)\gamma_0
\end{equation}
In evaluating the trace ${\rm{Tr}}(h\rho)$ we integrate over
x-coordinate, use orthonormality relations for $H_\nu(\zeta)$ and finally
using the anti-commutation relations for $ \gamma$-matrices, we have
the first term of the product $\propto 0$ (from the conclusion drawn
just after eqn.(21)), second term $\propto 4p_z^2
$, third term $\propto 4m^2$ and the fourth term $\propto
4p_\perp^2$.
Finally adding, we have ${\rm{Tr}}(\rho h)\propto 2E$ .
Since $E<0$, this is actually $\propto$  $-2E$, where
$E=(p_z^2+p_\perp^2+m^2)^{1/2}$
Then the kinetic energy density is given by 
\begin{eqnarray}
\epsilon_v^{(0)}&=&2N_c\sum_{f=u}^d\frac{q_fB_m}{2\pi^2}
\sum_{\nu=0}^{[\nu_{\rm{max}}^{(f)}]} (2-\delta_{\nu 0})
\int_0^\Lambda  dp_z E\nonumber\\ 
&=&\frac{N_c}{4\pi^2}\sum_{f=u}^dq_fB_m\sum_{\nu=0}^{[\nu_{\rm{max}}^{(f)}]}
(2-\delta_{\nu 0})
\Big[\Lambda(\Lambda^2+m_{\nu,f}^2)^{1/2}\nonumber\\ &+&
m_{\nu,f}^2\ln \mid
\frac{\Lambda+(\Lambda^2+m_{\nu,f}^2)^{1/2}}{m_{\nu,f}}
\mid\Big]
\end{eqnarray}
where $m_{\nu,f}^2=m^2+2\nu q_fB_m$ and $e_u=2/3 e$, $e_d=1/3e$ and $N_c=3$

\noindent {\bf{Interaction Energy (Direct):}}

It is very easy to show that ${\rm{Tr}}(\beta\gamma_5\rho)=0$:
\begin{equation}
\beta\gamma_5\rho=\gamma_x\gamma_y\gamma_z\frac{1}{2E}[EA+p_zA
\gamma_z+mA-p_\perp B\gamma_y]
\end{equation}
Then it is very easy to check that the above result follows from here.

To calculate the direct  term of interaction we have to evaluate the
${\rm{Tr}}(\beta\rho)$ integrated over $x$-coordinate, i.e.,
\begin{equation}
\int_{-\infty}^{+\infty}{\rm{Tr}}(\beta\rho)dx=\frac{1}{2E}
\int_{-\infty}^{+\infty}
[E\beta A+p_zA\gamma_z+mA-p_\perp B\gamma_y]dx
\end{equation}
since first, second and fourth terms  contain odd (single) number
$\gamma$ with $A$ the
contribution of those terms are zero. Only third term contributes to the
integral and is given by
\begin{equation}
\frac{m}{2E}\int_{-\infty}^{+\infty}{\rm{Tr}}(A)dx
=\frac{m}{2E}\int_{-\infty}^{+\infty}2[I_nu^2(x)+I_{\nu-1}^2(x)]dx
\end{equation}
using orthonormality relation for Hermite polynomials the above integral
reduces to $2m/E$. Since there is an identical expression with
integral over $x^\prime$, finally we get
\begin{equation}
V_{\rm{dir}}=-4gm^2[{\cal{V}}(\Lambda, m)]^2
\end{equation}

\newpage
\noindent {\bf{Momentum Integral:-}} 

\noindent {\bf{First Term:}}

\begin{eqnarray}
&&\left(\frac{N_c}{2\pi^2}\sum_{f=u}^d
q_fB_m\sum_{\nu=0}^{[\nu_{\rm{max}}^{(f)}]}
(2-\delta_{\nu 0})
\int_0^\Lambda dp_{1z}\right)\times\nonumber\\ &&
\left(\frac{N_c}{2\pi^2}\sum_{f=u}^d
q_fB_m\sum_{\nu=0}^{[\nu_{\rm{max}}^{(f)}]}
(2-\delta_{\nu 0})
\int_0^\Lambda dp_{2z}\right)\nonumber\\ 
&=&
\left(\frac{N_c}{2\pi^2}\sum_{f=u}^d
q_fB_m\sum_{\nu=0}^{[\nu_{\rm{max}}^{(f)}]} (2-\delta_{\nu 0})\Lambda\right)^2
\end{eqnarray}

\noindent {\bf{Second Term:}}

\begin{eqnarray}
&&\left(\frac{N_c}{2\pi^2}\sum_{f=u}^d
q_fB_m\sum_{\nu=0}^{[\nu_{\rm{max}}^{(f)}]}
(2-\delta_{\nu 0})
\int_0^\Lambda\frac{p_{1z} dp_{1z}}{E_1}\right)\times\nonumber\\ &&
\left(\frac{N_c}{2\pi^2}\sum_{f=u}^d
q_fB_m\sum_{\nu=0}^{[\nu_{\rm{max}}^{(f)}]}
(2-\delta_{\nu 0})
\int_0^\Lambda\frac{p_2z dp_{2z}}{E_2}\right)\nonumber\\ 
&=&
\left(\frac{N_c}{2\pi^2}\sum_{f=u}^d
q_fB_m\sum_{\nu=0}^{[\nu_{\rm{max}}^{(f)}]} (2-\delta_{\nu 0})
(\Lambda^2+m_{\nu,f}^2)^{1/2}m_{\nu,f}\right)^2
\end{eqnarray}

{\noindent {\bf{Third Term:}}

\begin{eqnarray}
&& m^2\left(\frac{N_c}{2\pi^2}\sum_{f=u}^d
q_fB_m\sum_{\nu=0}^{[\nu_{\rm{max}}^{(f)}]}
(2-\delta_{\nu 0})
\int_0^\Lambda\frac{ dp_{1z}}{E_1}\right)\times\nonumber\\ &&
\left(\frac{N_c}{2\pi^2}\sum_{f=u}^d
q_fB_m\sum_{\nu=0}^{[\nu_{\rm{max}}^{(f)}]}
(2-\delta_{\nu 0})
\int_0^\Lambda\frac{ dp_{2z}}{E_2}\right)\nonumber\\ 
&=&
m^2\left(\frac{N_c}{2\pi^2}\sum_{f=u}^d
q_fB_m\sum_{\nu=0}^{[\nu_{\rm{max}}^{(f)}]} (2-\delta_{\nu
0})\ln\left[\frac{\Lambda+(\Lambda^2+m_{\nu,f}^2)^{1/2}}{m_{\nu,f}}
\right]\right)^2
\end{eqnarray}

\noindent {\bf{Interaction Energy (Exchange Terms):}}
 
The product
\begin{eqnarray}
(\beta\gamma_5\rho_1)(\beta\gamma_5\rho_2) &=&
\gamma_1\gamma_2\gamma_3\frac{1}{2E_1}[E_1A_1+p_{1z}A_1\gamma_3+
mA_1-p_{1\perp}B_1\gamma_2]\times\nonumber\\
&&
\gamma_1\gamma_2\gamma_3\frac{1}{2E_2}[E_2A_2+p_{2z}A_2\gamma_3+
mA_2-p_{2\perp}B_2\gamma_2]
\end{eqnarray}
Now

\begin{eqnarray}
 &&\gamma_1\gamma_2\gamma_3\frac{1}{2E_1}[E_1A_1+p_{1z}A_1\gamma_3+
mA_1-p_{1\perp}B_1\gamma_2]\nonumber\\
&=& \frac{1}{2E_1}[E_1\gamma_1\gamma_2\gamma_3 A_1+p_{1z}\gamma_1\gamma_2
A_1+\gamma_1\gamma_2\gamma_3mA_1-p_{1\perp}\gamma_1\gamma_2\gamma_3B_1\gamma_2]
\end{eqnarray}
Hence
\begin{eqnarray}
(\beta\gamma_5\rho_1)(\beta\gamma_5\rho_2) 
&=& \frac{1}{4E_1 E_2}[E_1\gamma_1\gamma_2\gamma_3 A_1+p_{1z}\gamma_1\gamma_2
A_1+\gamma_1\gamma_2\gamma_3mA_1-p_{1\perp}\gamma_1\gamma_2\gamma_3B_1\gamma_2]\times\nonumber\\
&& [E_2\gamma_1\gamma_2\gamma_3 A_2+p_{2z}\gamma_1\gamma_2
A_2+\gamma_1\gamma_2\gamma_3mA_2-p_{2\perp}\gamma_1\gamma_2\gamma_3B_2\gamma_2]
\end{eqnarray}
Now we use the results
\begin{equation}
\gamma_1\gamma_2\gamma_3\gamma_1\gamma_2\gamma_3
=-
\gamma_1\gamma_2\gamma_3\gamma_1\gamma_3\gamma_2
=\gamma_1\gamma_2\gamma_1\gamma_2=-1,~~\gamma_1\gamma_2\gamma_1\gamma_2=-1,~~
\end{equation}
to obtain the ${\rm{Tr}}[(\beta\gamma_5\rho_1)(\beta\gamma_5\rho_2)]$,
which is given by
\begin{eqnarray}
{\rm{Tr}}[(\beta\gamma_5\rho_1)(\beta\gamma_5\rho_2)]&=&
\frac{1}{4E_1E_2}\left[-4E_1E_2-4p_{1z}p_{2z}-4m^2-4mE_2-4mE_1\right]
\nonumber\\ &=&
-\left[1+\frac{p_{1z}p_{2z}}{E_1E_2}+\frac{m^2}{E_1E_2}+m\left(\frac{1}{E_1}+
\frac{1}{E_2}\right)\right]
\end{eqnarray}

The integration over $p_{1z}$ and $p_{2z}$ are done in the same manner
as are did for ${\rm{Tr}}[(\beta\rho_1)(\beta\rho_2)]$.

\end{document}